\documentclass[preprint,1p]{elsarticle}

\usepackage{amssymb}

\journal{Astroparticle Physics}

\begin{document}

\begin{frontmatter}



\title{Galactic Cosmic-Ray Induced Production of Lithium in the Small Magellanic Cloud}


\author{A. \'Ciprijanovi\'c$^1$}

\address{$^1$Department of Astronomy, Faculty of Mathematics,
University of Belgrade\break Studentski trg 16, 11000 Belgrade,
Serbia, aleksandra@matf.bg.ac.rs}

\begin{abstract}
Recently, the first lithium detection outside of the Milky Way was made in low-metallicity gas of the Small Magellanic Cloud, which was at the level of the expected primordial value. Part of the observed lithium in any environment has primordial origin, but there is always some post-BBN (Big Bang Nucleosynthesis) contamination, since lithium can also be produced in cosmic-ray interactions with the interstellar medium. Using the fact that processes involving cosmic rays produce lithium, but also gamma rays through neutral pion decay, we use the Small Magellanic Cloud gamma-ray observations by \emph{Fermi}-LAT to make predictions on the amount of lithium in this galaxy that was produced by galactic cosmic rays accelerated in supernova remnants. By including both fusion processes, as well as spallation of heavier nuclei, we find that galactic cosmic rays could produce a very small amount of lithium. In the case of ${}^6\mathrm{Li}$ isotope (which should only be produced by cosmic rays) we can only explain $0.16\%$ of the measured abundance. If these cosmic rays are indeed responsible for such small lithium production, observed abundances could be the result of some other sources, which are discussed in the paper.
\end{abstract}


\end{frontmatter}


\section{Introduction}
\label{introduction}

Products of interactions of hadronic cosmic rays (CR) with the interstellar medium (ISM) can be used to probe the history as well as present day cosmic-ray production and interactions. High energy CRs produce gamma rays ($\mathrm{p}+\mathrm{p}\rightarrow \pi^{0}\rightarrow\gamma+\gamma$) through neutral pion decay~\citep{STE70,STE71}. Cosmic-ray collisions with the ISM can also produce light elements (lithium, beryllium and boron). For example, production of lithium can be the result of fusion ($\alpha+\alpha \rightarrow {}^{6,7}\mathrm{Li}$;~\cite{MO77}) and spallation of heavier nuclei ($\mathrm{p},\alpha+\mathrm{C,N,O}\rightarrow{}^{6,7}\mathrm{Li}$;~\cite{RV84}). Since both processes are the result of hadronic CR interactions they can be linked, as was done by Fields \& Prodanovi\'c (2005)~\cite{FP05}, who gave a simple model-independent
connection between lithium produced in fusion reactions and pionic gamma rays, produced by some cosmic-ray population. They linked Solar lithium abundances~\citep{AG89} and the isotropic diffuse gamma-ray background (which excludes any resolved sources) - IGRB~\citep{SB98} observed by the EGRET telescope~\citep{FP05, PF05}. This relation can be used to constrain any CR population from two sides. Still, the difference between these two CR products, does exist. Production of lithium is a cumulative process, and the present day abundances of lithium in the gas of any galaxy are the result of CR production over the history of the galaxy (plus some primordial, as well as some pre-galactic production). On the other hand, gamma-ray luminosity of a galaxy at any given moment is the result of the CR interactions at that moment.

First measurements of ${}^7\mathrm{Li}$ outside of the Milky Way (MW) were made in the low-metallicity interstellar gas of the Small Magellanic Cloud (SMC), which has one quarter of the Sun's metallicity~\citep{HL12}. Observation of ${}^7\mathrm{Li}$ in low-metallicity interstellar gas is important since these abundances should not be affected by changes in stellar atmospheres that could be present in lithium abundances measured in for example MW halo stars. So, in case of lithium in the interstellar gas it might be easier to distinguish between primordial and CR-produced components of lithium. The measured value $({}^7\mathrm{Li}/\mathrm{H})_\mathrm{SMC,obs}=(4.8\pm1.8)\times10^{-10}$ ~\citep{HL12} is at the level of the expected primordial abundance $({}^7\mathrm{Li}/\mathrm{H})_\mathrm{BBN}=(4.56-5.34)\times10^{-10}$~\citep{CUV14}. Measurements also give the isotopic ratio $({}^6\mathrm{Li}/{}^7\mathrm{Li})_\mathrm{SMC,obs}=0.13\pm0.05$, with the formal limit of $({}^6\mathrm{Li}/{}^7\mathrm{Li})_\mathrm{SMC,obs}<0.28\,(3\sigma)$~\citep{HL12}.

This small neighboring galaxy is also interesting since it was detected in gamma rays by \emph{Fermi} telescope with an integrated flux of $F_{\gamma,\mathrm{SMC,obs}}(>100\,\mathrm{MeV})=(3.7\pm0.7)\times10^{-8}\,\mathrm{phot}\,\mathrm{cm}^{-2}\mathrm{s}^{-1}$~\citep{AA10}. Here we will use model-independent gamma-ray--$\mathrm{Li}$ connection~\citep{FP05,PF05} and models for production of GCRs in normal galaxies~\citep{PF01,PF02} to constrain the post-BBN lithium production by GCRs in SMC and check weather GCRs could produce an important part of the observed SMC lithium abundance.


\section{Formalism and Results}
\label{formalismandresults}

The connection between CR lithium synthesis and hadronic gamma-ray production is derived in Fields \& Prodanovi\'c (2005)~\cite{FP05}. Low-energy ($\approx10-70\,\mathrm{MeV}\,\mathrm{nucleon}^{-1}$) cosmic rays produce lithium through $\alpha\alpha\rightarrow{}^{6,7}\mathrm{Li}+...$, but cosmic rays also produce gamma rays via neutral pion decay $\mathrm{pp}\rightarrow\pi^0\rightarrow\gamma\gamma$ (threshold energy for this process is higher $\approx280\,\mathrm{MeV}\,\mathrm{nucleon}^{-1}$).

We want to use this connection to link the production of lithium in SMC and the pionic gamma-ray intensity produced by all SMC-like galaxies over the cosmic history. We will assume that both lithium and gamma rays are produced inside galaxies, and that the cumulative pionic gamma-ray intensity produced by all SMC-like galaxies is isotropic, even though it is produced as the sum of all unresolved small galaxies, which are point sources. In~\cite{FP05} the relation linking these two quantities - pionic gamma-ray intensity $I_{\gamma}$ (integrated over the entire energy range) and lithium abundance, specifically mole fraction $Y_{6,7} \equiv \frac{n_{6,7}}{n_\mathrm{b}}$ (where $n_{6,7}$ is number density of ${}^{6,7}\mathrm{Li}$, and $n_\mathrm{b}$ comoving baryon number density), produced in fusion reactions of CRs and $\alpha$ particles from the ISM was given as:

\begin{equation}
\frac{I_{\gamma}(E>0,t)}{Y_{6,7}(\overrightarrow{x},t)}=\frac{n_{\mathrm{b}}c}{4\pi y_{\alpha}^{\mathrm{cr}} y_{\alpha}^{\mathrm{ism}}} \frac{\sigma_\gamma}{\sigma^{6,7}_{\alpha\alpha}}\frac{F_\mathrm{p,avg}(t)}{F_\mathrm{p}(\overrightarrow{x},t)},
\end{equation}
where the comoving baryon number density is $n_\mathrm{b}=2.52\times10^{-7}\,\mathrm{cm}^{-3}$ (and enters equation (1) because of the way~\cite{FP05} define mole fraction $Y_{6,7}$ and pionic gamma-ray intensity $I_{\gamma}$). The CR and ISM helium abundances are $y_{\alpha}^{\mathrm{cr}}=y_{\alpha}^{\mathrm{ism}}=0.072$ where $y_\alpha= n_\alpha/n_\mathrm{H}\equiv(\mathrm{He}/\mathrm{H
})$ and $n_\mathrm{H}$ is hydrogen number density (based on SMC abundances from~\cite{RV03}). The flux-averaged pionic gamma-ray production cross section is $\sigma_{\gamma} \equiv 2\xi_{\alpha} \zeta_{\pi} \sigma_{\pi^0}$, where
the factor of $2$ is introduced because of the number of photons per pion decay, the factor $\xi_{\alpha}=1.45$ accounts for $\mathrm{p}\alpha$ and $\alpha\alpha$ reactions~\citep{DE86}, $\zeta_{\pi}$ is the pion multiplicity and $\sigma_{\pi^0}$ is the flux-averaged cross section for pion production. For the total inclusive cross section $\zeta_{\pi}\sigma_{\pi^0}$ for pion production we use parametrization from~\citep{NO09}. Cross sections for lithium production $\sigma^{6,7}_{\alpha\alpha}$ is from~\citep{MA01}. We will focus on ${}^6\mathrm{Li}$ isotope since it is only produced in CR interactions with the ISM~\citep{FO99,VF99}, so the measured abundances of this lithium isotope can tell us more about the CR flux in that environment.

Lithium and gamma-ray production cross sections behave differently, since lithium production is a low-energy phenomenon, while neutral pion production is a significantly higher energy phenomenon. These differences are sensitive to the choice of cosmic-ray spectra one uses. The best fit of the observed SMC gamma-ray spectrum from~\cite{AA10} gives a power law spectrum in total energy, with spectral index $\alpha_\mathrm{SMC}=2.23$ (in the $0.1-500\,\mathrm{GeV}$ energy range).  We will adopt a source spectrum that is a power law in momentum $\phi(p)\propto p^{-2.23}$, and include cosmic-ray propagation based on ''closed box'' model. Propagation was implemented like in~\cite{FO94}, but without the particle escape (which was included in their paper for calculating light element abundances in the Milky Way, which can be described using ''leaky box'' model). Propagation will mostly impact lower energies, which are the most important when considering lithium production (so the propagated spectrum in the $\mathrm{MeV}$ energy range will be different than the approximation of a simple power law in total energy). On the other, ''closed box'' model doesn't allow for particle escape, so the slope of the propagated spectrum at higher energies will not change significantly and will remain consistent with the observed one at energies over a few $\mathrm{GeV}$. Cross sections for gamma-ray and lithium production used in Eq.(1) are averaged over this propagated CR spectrum, where integration is done from threshold energy $E_\mathrm{th}=3\,\mathrm{MeV}/\mathrm{nucleon}$.\footnote{This is the lowest threshold energy for all reactions considered here, i.e. $280\,\mathrm{MeV}$ in case of pionic gamma-ray production~\citep{NO09}, $8.75\,\mathrm{MeV}/\mathrm{nucleon}$ in case od fusion~\citep{MA01} and $3\,\mathrm{MeV}/\mathrm{nucleon}$ in case of spallation production of lithium~\citep{RV84} (this lowest threshold energy is for $\alpha+\mathrm{N}\longrightarrow{}^6\mathrm{Li}+...$ reaction).} After averaging, the ratios of average cross sections for pion and $\alpha\alpha$ lithium production are $\sigma^{6}_{\alpha\alpha}/\sigma_{\pi^0}=0.72$ and $\sigma^{7}_{\alpha\alpha}/\sigma_{\pi^0}=1.28$. The ratio of average cross sections for production of the two lithium isotopes in case of this CR spectrum is $\sigma^{7}_{\alpha\alpha}/\sigma^{6}_{\alpha\alpha}=1.78$. Still, extrapolation to lower energies we are interested in, is always uncertain, and is based on assumptions of CRs propagation. Resent observations of local interstellar spectrum, which goes as low as $5\,\mathrm{MeV}$, suggests the presence of more low energy CRs, than what would be expected from standard propagated galactic CR spectrum~\citep{P14}, like the one we use.

The ratio of the line-of-sight baryon-averaged cosmic-ray fluence (time-integrated flux) and local CR fluence in some star-forming galaxy (weighed by its gas mass fraction $\mu(\overrightarrow{x},t)$) ${F_\mathrm{p,avg}(t)}/{F_\mathrm{p}(\overrightarrow{x},t)}$ compares the cumulative CR activity in an average-star forming galaxy and local CR activity in the considered galaxy~\citep{FP05}. In this paper we will compare the mean cosmic-ray fluence inside small irregular SMC-like galaxies and local CR fluence inside the SMC itself. If we assume that SMC is a typical representative of these small galaxies (with masses similar to that of the SMC) the fluence ratio will be ${F_\mathrm{p,avg}(t)}/{F_\mathrm{p}(\overrightarrow{x},t)}\approx1$.

Gamma-ray observations of SMC by \emph{Fermi}-LAT have resulted in an integrated gamma-ray flux $F_{\gamma,\mathrm{SMC,obs}}(>100\,\mathrm{MeV})=(3.7\pm0.7)\times10^{-8}\,\mathrm{phot}\,\mathrm{cm}^{-2}\mathrm{s}^{-1}$~\citep{AA10}. This flux is the result of cosmic-ray activity of SMC at the present epoch. On the other hand non-primordial part of the measured lithium abundance in the SMC was produced by cosmic rays throughout SMC's history. Here we will use the present day gama-ray flux of the SMC measured by the \emph{Fermi}-LAT to normalize the differential gamma-ray intensity $\frac{\mathrm{d}I_\gamma}{\mathrm{d}\Omega\,}\left[\mathrm{phot}\,\mathrm{GeV}^{-1}\mathrm{cm}^{-2}\mathrm{s}^{-1}\mathrm{sr}^{-1}\right]$, detected at the present epoch but produced by a population of SMC-like galaxies that have accelerated cosmic rays over the cosmic time (based on models from~\cite{PF02}):

\begin{eqnarray}
&&\frac{\mathrm{d}I_{\gamma}}{\mathrm{d}\Omega}=\frac{c}{4\pi H_0 \psi_\mathrm{SMC}}\int_0^{z_\ast}\mathrm{d}z\left[\dot{\rho}_{*}(z)\frac{C\Gamma(E(1+z),z)}{\sqrt{\Omega_\Lambda+\Omega_\mathrm{m}(1+z)^3}}\right.\\\nonumber
&&\times\left.\left(\frac{1}{\mu_{0,SMC}}-\left(\frac{1}{\mu_{0,SMC}}-1\right)\right)\frac{\int_{z_\ast}^z \mathrm{d}z\frac{\mathrm{d}t}{\mathrm{d}z}\dot{\rho_{*}}(z)}{\int_{z_\ast}^0 \mathrm{d}z\frac{\mathrm{d}t}{\mathrm{d}z}\dot{\rho_{*}}(z)}\right]\,.\\\nonumber
\end{eqnarray}

We use $H_0=67.81\,\mathrm{km}\mathrm{s}^{-1}\mathrm{Mpc}^{-1}$ as a Hubble parameter, and $\Omega_\Lambda=0.692$ and $\Omega_\mathrm{m}=0.308$ as values for cosmological parameters, which are all based on Planck measurements~\citep{Ade15}. For the SMC gas mass fraction today we adopt $\mu_{0,\mathrm{SMC}}=M_\mathrm{gas,SMC}/M_\mathrm{tot,SMC}=0.11$, which was derived using total SMC mass of $M_\mathrm{tot,SMC}\approx4\times10^9\,M_\odot$~\citep{HZ06} and gas mass $M_\mathrm{gas,SMC}=4.5\times10^8\,M_\odot$~\citep{SS99}. Present star formation rate of the SMC is $\psi_{\mathrm{SMC}}=0.3\,M_\odot\,\mathrm{yr}^{-1}$~\citep{RJ14}.

The evolution of GCR flux over cosmic time is imprinted in the evolution of cosmic star formation rate (CSFR) $\dot{\rho}_{*}(z)\,\left[M_\odot\mathrm{yr}^{-1}\mathrm{Mpc}^{-3}\right]$, for which we use results from the Illustris simulation~\citep{VG14}. Their total CSFR is in quite good agreement with the observations, and only slightly overestimates total CSFR at lowest redshifts (this might be due lower AGN feedback, which can result in larger star formation at these redshifts). In \cite{VG14} they also give CSFR curves for galaxies in different stellar mass bins in the range of $10^{7}-10^{11}\,M_\odot$. In case of the SMC with stellar mass $M_\mathrm{star,SMC}=4\times10^{8}\,M_\odot$~\citep{RJ14} we will adopt the closest CSFR curve, which corresponds to objects with stellar masses of $10^{8.5}\,M_\odot$ and which describes star formation in galaxies similar to the SMC. Smaller galaxies, have a much more important contribution to the total CSFR on larger redshifts, while larger galaxies such as the MW have mostly virialized around $z_\ast=5$ and dominate star formation at low redshifts. Based on the CSFR curve for small SMC-like galaxies, for the redshift of virialization of these galaxies we will adopt $z_\ast=10$.

For the shape of the gamma-ray spectrum produced by neutral pion decay (which were formed by GCR interactions with the ISM in the galaxy) $\Gamma(E,z)$, we use semi-analytical formula from~\cite{PE03} and we normalize it using the observed gama-ray flux of the SMC. Finally using~\cite{PF02} GCR models and input parameters described here, we find the resulting differential gamma-ray intensity of the SMC-like galaxies and plot it on Fig. 1 (solid line). We also plot the latest \emph{Fermi}-LAT diffuse gamma-ray background measurements~\citep{AA15}. Based on our model SMC-like galaxies contribute very little to the diffuse gamma-ray background (around $0.15\%$). We did not include attenuation of high energy gamma rays by the extragalactic background light~\citep{SS98,GM09,GS12}, since this type of attenuation is significant at energies $>100\,\mathrm{GeV}$, which are higher than energies important for our study.

The resulting differential gamma-ray intensity describes the entire cosmic-ray activity of SMC-like galaxies over the cosmic time. It gives the total gamma-ray intensity of $I_{\gamma}(>0)=3.08\times10^{-8}\mathrm{phot}\,\mathrm{cm}^{-2}\mathrm{s}^{-1}\mathrm{sr}^{-1}$ which can now be used in Eq.(1) to find the resulting ${}^6\mathrm{Li}$ abundance. The abundance of this lithium isotope that could be produced by GCRs (if we assume that the whole SMC gamma-ray flux observed by the \emph{Fermi}-LAT is produced by those GCRs) is:
$$({}^6\mathrm{Li}/\mathrm{H})_\mathrm{SMC}=8.69\times10^{-14}\,.$$
If we adopt the isotopic ratio of $({}^6\mathrm{Li}/{}^7\mathrm{Li})=0.13$ in the SMC than the corresponding observed ${}^6\mathrm{Li}$ abundance is $({}^6\mathrm{Li}/\mathrm{H})_\mathrm{SMC,obs}=6.24\times10^{-11}$. In this case the lithium abundance derived here represents $0.14\%$ of the observed value.

\begin{figure}
\centerline{\includegraphics[width=1\columnwidth, keepaspectratio]{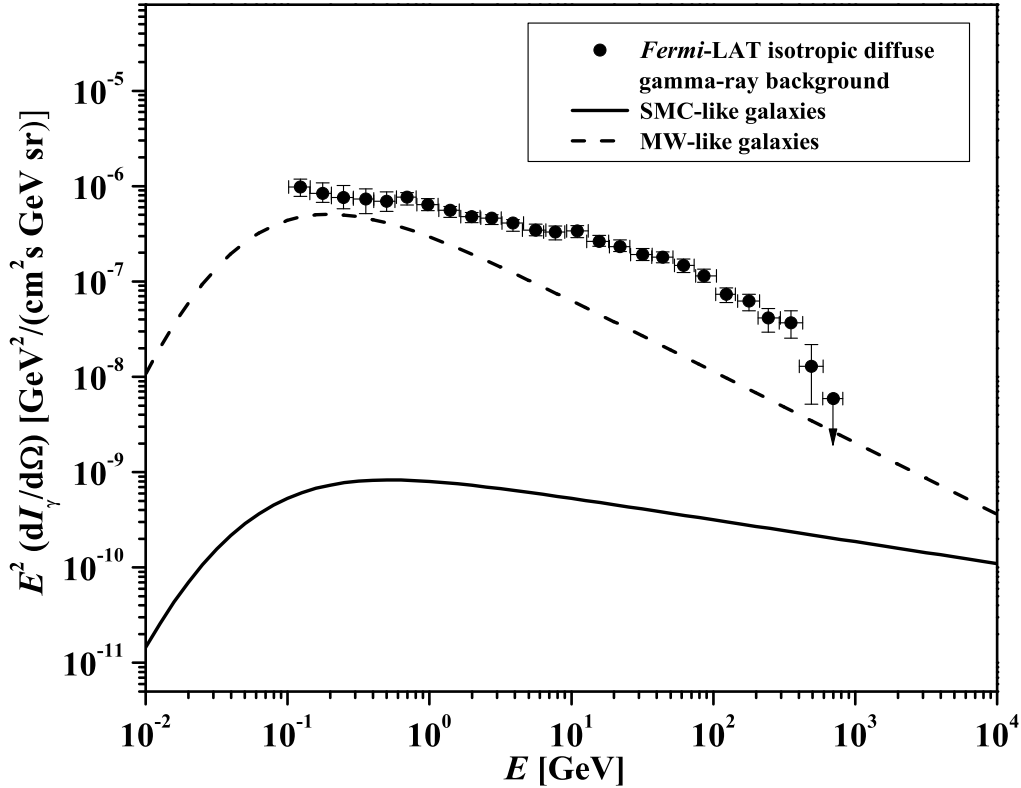}}
\caption{Differential gamma-ray intensity of SMC-like galaxies (solid line). Data points represent the latest \emph{Fermi}-LAT measurements of the isotropic diffuse gamma-ray background (IGRB)~\citep{AA15}. For comparison, we also plot differential gamma-ray intensity of MW-like galaxies (dashed line) derived in the same way.}
\end{figure}

Fusion reactions are not the only processes by which CRs produce lithium. Spallation of heavier nuclei (C,N,O) in the ISM, by protons and alpha particle from CRs, can also produce lithium via $\mathrm{p},\alpha+\mathrm{C,N,O}\rightarrow{}^{6,7}\mathrm{Li}$. Reverse processes in which heavier nuclei CRs interact with $\mathrm{p}$ and $\alpha$ in the surrounding ISM, also lead to the same reaction. In case of fusion processes the produced lithium mole fraction $Y^{\alpha\alpha}_6\sim y_\alpha^\mathrm{cr}Y_\alpha^\mathrm{ism}\sigma^{\alpha\alpha}_6\Phi_\mathrm{p}t_0$ depends on the abundances of $\alpha$ particles in the ISM $Y_\alpha^\mathrm{ism}$, flux of CR alpha particles which can be expressed using the flux of CR protons as $y_\alpha^\mathrm{cr}\Phi_\mathrm{p}$, production cross section $\sigma^{\alpha\alpha}_6$, and the age of the galaxy considered $t_0$. Similar relation can be written in case of lithium produced in spallation processes. We will include both forward and inverse kinematics spallation reactions. Abundances of C,N and O in the SMC can be found in~\cite{RV03}, and we have used abundances derived from measurements from main-sequence stars and HII regions. We approximate here that SMC abundances were constant through SMCs history and equal to the half of the present day value. In case of our closed box propagated spectrum and SMC abundances we get that spallation produced mole fraction of ${}^6\mathrm{Li}$ is $Y^\mathrm{spall}_6=0.16Y^\mathrm{\alpha\alpha}_6$. So after including spallation the total amount of lithium that can be produced by GCRs is $$({}^6\mathrm{Li}/\mathrm{H})^\mathrm{tot}_\mathrm{SMC}=1.0\times10^{-13}\,,$$
which represents only $0.16\%$ of the abundance of ${}^6\mathrm{Li}$ measured in the SMC gas.

The same calculation can be applied in any case where lithium and gamma-ray data are available. For example,~\cite{FP05} have used the same gamm-ray--lithium connection in case of the Milky Way (MW). As a consistency check, we will also use equation (1) to link MW lithium abundances and the gamma-ray intensity of all unresolved MW-like galaxies. In case of the MW we will use present star formation rate $\psi_\mathrm{MW}=1.9\,M_\odot\,\mathrm{yr}^{-1}$~\citep{CP11}, gas mass fraction $\mu_{0,\mathrm{MW}}=0.12$ derived with total mass of our Galaxy $M_\mathrm{tot,MW}=8\times10^{10}\,M_\odot$ and gas mass $M_\mathrm{gas,MW}=1\times10^{10}\,M_\odot$, and gamma-ray luminosity $L_\mathrm{MW}(>100\,\mathrm{MeV})=2.86\times10^{42}\,\mathrm{phot}\,\mathrm{s}^{-1}$, all from~\cite{PF02}. For the CSFR curve we will use Illustris curve for total CSFR~\cite{VG14}, and since larger MW type galaxies are the main source of star formation on lower redshifts, we can still approximate ${F_\mathrm{p,avg}(t)}/{F_\mathrm{p}(\overrightarrow{x},t)}\approx1$. In case of the MW we will use the same momentum power law source spectrum $\phi(p)\propto p^{-2.23}$ (consistent with~\cite{SW14} for local GCR spectrum), but the propagation is done based on ''leaky box'' model just as in~\cite{FO94}, which allows the escape of the most energetic particles and leads to the change of spectral index on high energies (where we get $\alpha\approx2.75$). Abundances of C,N,O that we need in order to include spallation processes can be found in~\cite{AG89} (we again assume that these abundances were constant throughout MW history and equal to the half of the present day value). The resulting differential gamma-ray intensity of MW-like galaxies is also plotted on Fig. 1 (dashed line). It is in good agreement with the~\cite{FPP10} results for the total GCR-produced gamma-ray spectrum of all unresolved star forming galaxies (they have also assumed that MW is a typical star forming galaxy). We calculate the possible ${}^6\mathrm{Li}$ production by GCRs in the MW, and get that MW could produce around $3$ orders of magnitude more lithium than SMC. Compared to the solar abundance of lithium $({}^6\mathrm{Li}/\mathrm{H})_\odot=1.53\times10^{-10}$~\citep{AG89} we get $({}^6\mathrm{Li}/\mathrm{H})_\mathrm{MW}\approx0.76({}^6\mathrm{Li}/\mathrm{H})_\odot$. This also means that if we assume that the entire solar ${}^6\mathrm{Li}$ abundance is produced by GCRs, the corresponding gamma-ray intensity of the MW-like galaxies would be larger and would violate the observed IGRB, which is consistent with the conclusions in~\cite{FP05,PF06}.

The difference in resulting lithium production when using SMC and MW for normalization comes from a larger CSFR in case of larger MW galaxy. Also, the present star formation rate of the MW $\psi_\mathrm{MW}=1.9\,M_\odot\,\mathrm{yr}^{-1}$~\citep{CP11} is around $6$ times larger than in the SMC $\psi_{\mathrm{SMC}}=0.3\,M_\odot\,\mathrm{yr}^{-1}$~\citep{RJ14}. Our galaxy also has over an order of magnitude larger gas mass compared to the SMC, which results in larger cosmic-ray lithium production and the resulting gamma-ray flux. There is also a difference in the slope of the CR spectra for these two galaxies, which will impact the mean production cross sections for lithium and gamma-rays. Moreover, in~\cite{AA10} gamma-ray emissivity of the SMC is compared to that of the MW and they argue that the local MW value is at least $6-7$ times higher then that of the SMC, while~\cite{PF01,PF02} use an even higher value for the MW gamma-ray emissivity. Also, in~\cite{AA10} it is shown that the observed gamma-ray flux in the SMC implies that the average CR density in this galaxy is at most around $15\%$ of the measured MW value. All of this will result in a smaller lithium production in the SMC compared to our Galaxy.


\section{Discussion}
\label{discussion}

Using the \emph{Fermi}-LAT gamma-ray detection of the SMC we have estimated the amount of lithium that can be produced by GCRs which we assume to be the cosmic-ray population producing observed gamma-ray flux. Even though, GCRs are expected to be the dominant CR population, our calculation shows that GCR-produced ${}^6\mathrm{Li}$ is less then $1\%$ of the observed abundance. We have included $\alpha\alpha$ interactions, as well as spallation processes. In case of SMCs metallicity and CR spectrum, spallation is a subdominant production channel and most of the CR produced lithium in case of this galaxy is produced via fusion processes.

As the most extreme assumption, we can say that the \emph{Fermi}-LAT diffuse gamma-ray background~\citep{AA15} is entirely produced by gamma rays from unresolved SMC-like galaxies. If we go through the same procedure as before, with this extreme assumption, we can explain $55\%$ of the observed ${}^6\mathrm{Li}$ abundance. On the other hand this extreme assumption can produce only $12\%$ of the observed ${}^7\mathrm{Li}$ abundance (and $0.04\%$ without this extreme assumption), which is consistent with the fact that this lithium isotope, is in big part produced in the BBN. Observed abundance of this isotope in the SMC is consistent with the expected primordial abundance, so after removing GCR-produced ${}^7\mathrm{Li}$ that we get, we still won't deviate much from the expected primordial abundance. Also, in~\cite{HL12} it was found that the observed lithium isotopic ratio of $0.13$ implies that CRs in general could have produced $19\%$ of the observed ${}^7\mathrm{Li}$ abundance in the SMC. All of this, leaves room for some additional CR component next to the GCRs.

We can also use the observed lithium abundances in the SMC and predict how much can SMC-like galaxies contribute to the observed IGRB if the entire observed lithium in SMC was produce solely by GCRs. If this was the case, the resulting gamma-ray production would be $1.8$ times larger than the observed IGRB in case of the propagated CR spectrum (if we include both fusion and spallation processes). This would also mean that the present day gamma-ray flux of the SMC should be $3$ orders of magnitude larger than the observed value. On the other hand, if an important part of the observed SMC lithium is made by some other process, other than GCR interactions, it would be possible to produce the observed lithium abundances without producing that many gamma-rays.

The isotopic ratio of lithium measured in the SMC is consistent with the ratio measured in the MW gas~\citep{KK09}. Since SMC is a lower metallicity system and with a lower CR fluence than in the MW, if GCRs are the dominant CR species one would expect the isotopic ratio ${}^6\mathrm{Li}/{}^7\mathrm{Li}$ to be lower than in the MW. Observed abundance of ${}^6\mathrm{Li}$, which is larger than the one derived from our models, and the observed isotopic ratio work in favor of some other cosmic-ray acceleration mechanism being present in the SMC, which could in turn produce more lithium. For example cosmic rays could be produced by large scale structure formation~\citep{MR00,FL04,FP05,DP14}, high-energy pulsars~\citep{OG69,AA10}, pulsar wind nebulae~\citep{BE07} or binary systems~\citep{RC05,DU13}. If any of these mechanisms is an important contributor to lithium production and accelerates CRs for a long time, it would also enhance the gamma-ray flux of these galaxies. The observed IGRB can constrain any of these mechanism, but since we are dealing with small galaxies (and our GCR produced gamma-ray intensity is much lower than the observed IGRB), there is still room for additional CR populations. Also, SMC might not be a typical small galaxy, corresponding to the CSFR curve we have used. For example SMC might have had periods of much larger star formation than the present one, triggered by close flybys with the MW and their tidal interaction~\citep{PB13,DP15}. This could also lead to the production of additional CR population, only present during galaxy interaction. This CR population would not be present for a long time compared to the life of a galaxy, so their gamma-ray production in all unresolved SMC-like galaxies would not violate the observed IGRB, nor would it affect the present day gamma-ray flux of the SMC. Also, these tidal interactions would impact smaller galaxies more, so more lithium would be produced in the SMC, but not in the MW. In~\cite{PB13} it was shown that tidal interactions could be a very effective mechanism for lithium production in the SMC, and that even a single close fly-by of MW and SMC could produce a non-negligible abundance of lithium. The observed gamma-ray flux and lithium abundances of the SMC imply a more interesting galactic history of this galaxy in order to fully explain both measurements and understand how the CR flux inside the galaxy has changed during the lifetime of the SMC, which is something we will consider in the follow-up work.

\bigskip
\emph{Acknowledgments} I am grateful to Tijana Prodanovi\'c for help and guidance during my work on this paper. This work was supported by the Ministry of Science of the Republic of Serbia under project number 176005.






\end{document}